\documentclass[preprint2]{aastex} 
\usepackage{amsmath}
\usepackage{epsfig}
\usepackage{color}

\def\figone{
\begin{figure*}[t]
\centering
\includegraphics[width=\textwidth, keepaspectratio=true]{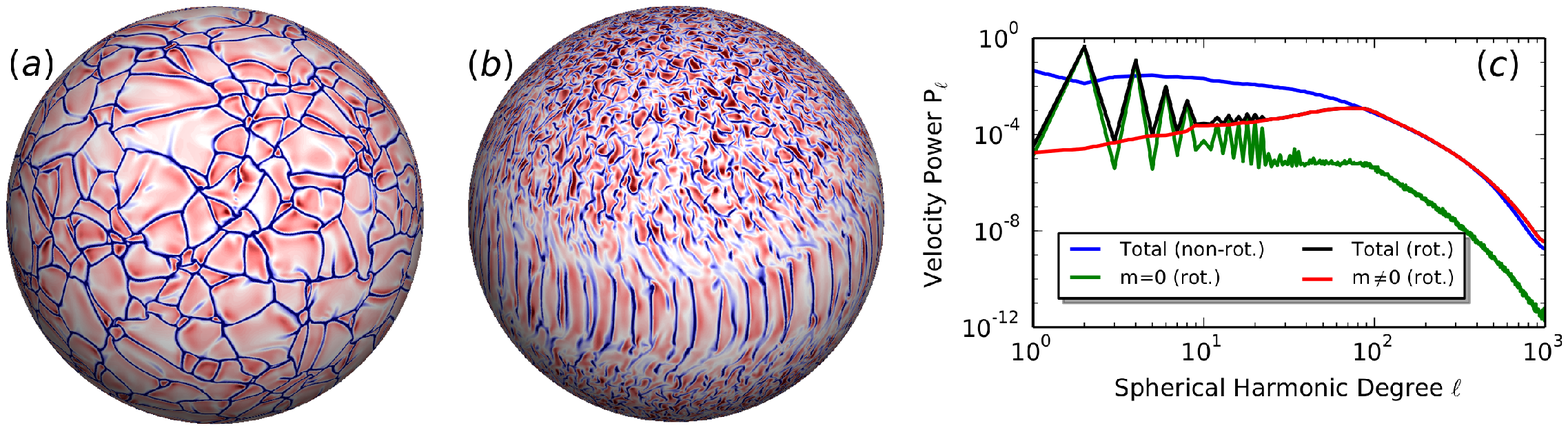}
\caption{\footnotesize Radial velocity $v_r$ as realized at Rayleigh number Ra$_\mathrm{F}$ = 6.81$\times$ 10$^{6}$ in (\textit{a}) the absence of rotation and in (\textit{b}) its presence, with Ekman number Ek = 1.91 $\times$ 10$^{-4}$.  Red (blue) tones denote upflows (downflows). Flows have been sampled at a radius $r/r_\mathrm{outer}= 0.99$.  Flow patterns in the rotating case possess markedly smaller spatial scales than their non-rotating counterparts. (\textit{c})  Horizontal velocity power spectra for the non-rotating model (\textit{a}; blue) and the rotating model (\textit{b}; black), normalized to have unit integrated power.  The power associated with axisymmetric flows ($m$=0; green) and convective flows ($m\neq0$; red) is shown for the rotating model.  Convective power possesses a prominent peak at $\ell=77$ in the rotating model.  }
\label{fig:composite}
\end{figure*}

}

\def\figtwo{
\begin{figure*}[t]
\centering
\includegraphics[width=\textwidth, keepaspectratio=true]{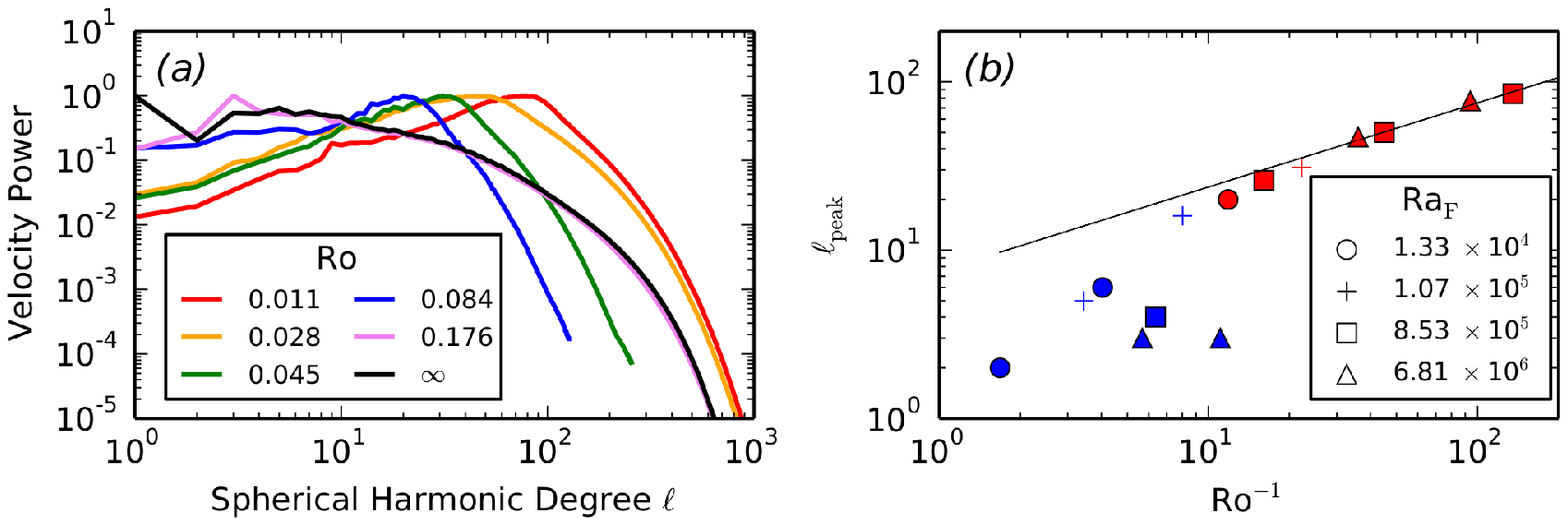}
\caption{\footnotesize (\textit{a})  Normalized spherical harmonic spectra of horizontal, convective velocity power as realized for a range of Rossby numbers Ro. As Ro decreases, spectral power peaks at higher $\ell$-values.  The violet curve corresponds to a model with anti-solar differential rotation; its spectrum resembles the non-rotating case (black curve).  All spectra have been time-averaged over (at least) one third of a viscous diffusion time and normalized to a peak value of one.  (\textit{b}) Wavenumber of maximum power $\ell_\mathrm{peak}$ for all models as a function of Ro.  Symbol shapes denote the value of Ra$_\mathrm{F}$.  Symbol colors denote solar-like (red) or anti-solar (blue) differential rotation. A line corresponding to $\ell_\mathrm{peak}\sim \mathrm{Ro}^{-1/2}$ is plotted for reference.}
\label{fig:progression}
\end{figure*}

}

\def\figthree{
\begin{figure}[t]
\centering
\includegraphics[trim = {7cm 3.5cm 6.5cm 3cm}, clip,width=0.95\columnwidth, keepaspectratio=true]{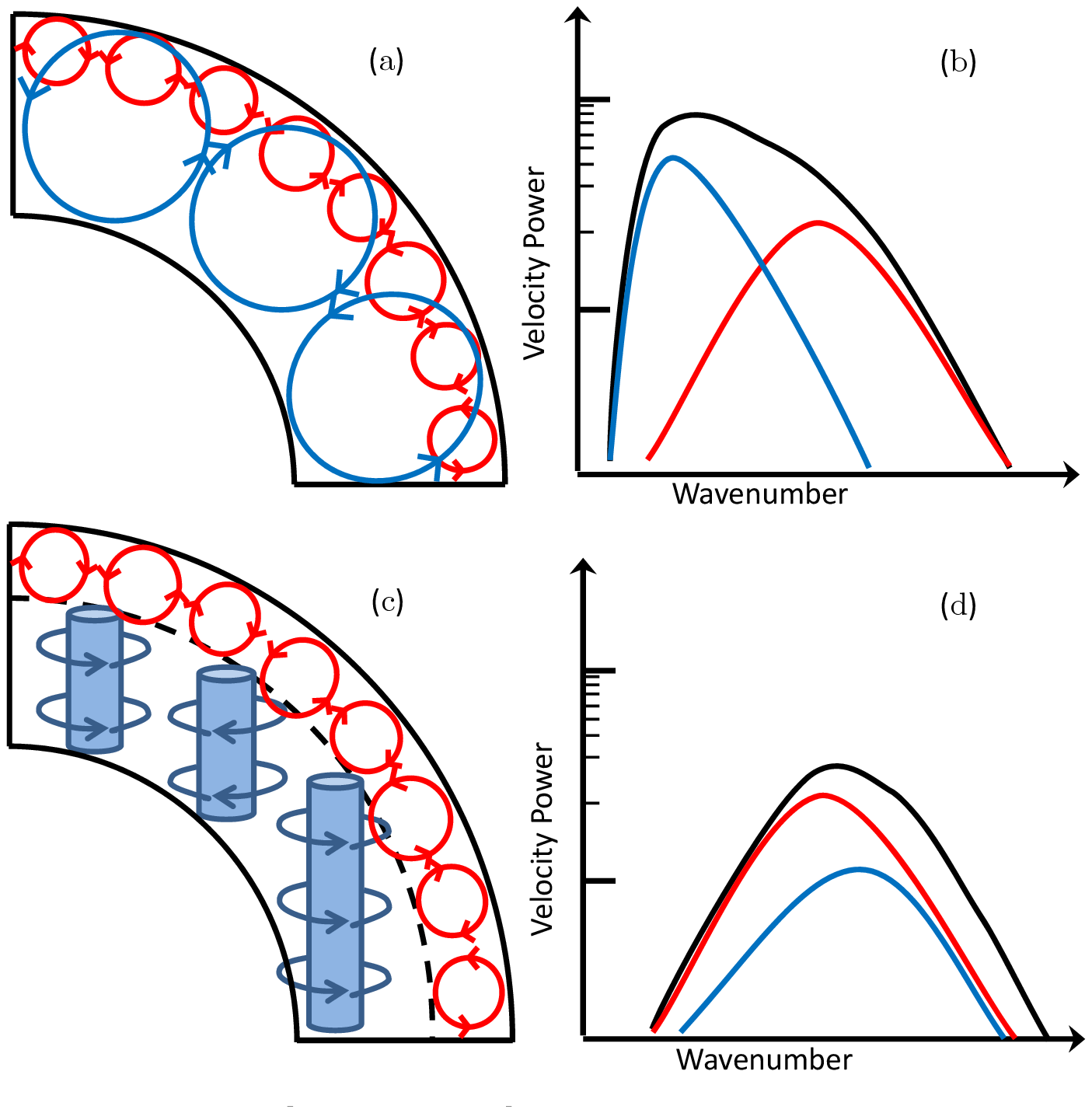}
\caption{\footnotesize Schematic of expected convective structures and their horizontal velocity spectra arising in (\textit{a},\textit{b}) non-rotating convection and in (\textit{c},\textit{d}) rotating convection.  In each instance, near-surface motions and their spectral contribution are depicted in red, deep-seated motions are indicated in blue, and their combined power is shown in black.  In the absence of rotation, convective power spans a range of spatial scales, peaking at the giant-cell scale (blue lines, panels $a$ and $b$).  When rotation is present, high-Ro convection manifests in the near-surface region (dashed line, panel $c$), and deep convection assumes a columnar, roll-like structure (blue cylinders, panel $c$).  Spectral power peaks at higher wavenumbers in the rotating system than in the non-rotating system.}
\label{fig:cartoon}
\end{figure}
}
\def\tableone{
\begin{table}[h]
\centering\small
\begin{tabular}[t]{rrr|rr}\\
\multicolumn{5}{c}{Table 1: Simulation Parameters}\\\hline
\multicolumn{3}{c}{Model Inputs} &  \multicolumn{2}{c}{Outputs}\\\hline
 \multicolumn{1}{c}{Ra$_\mathrm{F}$} & \multicolumn{1}{c}{Ek} & \multicolumn{1}{c}{$\ell_{\mathrm{max}}$} & \multicolumn{1}{c}{Ro} & \multicolumn{1}{c}{$\ell_{\mathrm{peak}}$}
\\\hline\hline

1.33 $\times$ 10$^4$ & 1.22 $\times$ 10$^{-2}$ & 127 & 0.084 & 20\\
1.33 $\times$ 10$^4$ & 2.45 $\times$ 10$^{-2}$ & 127 & 0.247 & 6\\
1.33 $\times$ 10$^4$ & 4.89 $\times$ 10$^{-2}$ & 127 & 0.595 & 2\\
1.07 $\times$ 10$^5$ & 3.06 $\times$ 10$^{-3}$ & 255 & 0.045 & 31\\
1.07 $\times$ 10$^5$ & 6.12 $\times$ 10$^{-3}$ & 255 & 0.125 & 16\\
1.07 $\times$ 10$^5$ & 1.22 $\times$ 10$^{-2}$ & 255 & 0.291 & 5\\

8.53 $\times$ 10$^5$ & 3.82 $\times$ 10$^{-4}$ & 511 & 0.007 & 85\\
8.53 $\times$ 10$^5$ & 7.65 $\times$ 10$^{-4}$ & 511 & 0.022 & 50\\
8.53 $\times$ 10$^5$ & 1.53 $\times$ 10$^{-3}$ & 511 & 0.062 & 26\\
8.53 $\times$ 10$^5$ & 3.06 $\times$ 10$^{-3}$ & 511 & 0.157 & 4\\

6.81 $\times$ 10$^6$ & 1.91 $\times$ 10$^{-4}$ & 1023 & 0.011 & 71\\
6.81 $\times$ 10$^6$ & 3.82 $\times$ 10$^{-4}$ & 1023 & 0.028 & 43\\
6.81 $\times$ 10$^6$ & 7.65 $\times$ 10$^{-4}$ & 1023 & 0.090 & 3\\
6.81 $\times$ 10$^6$ & 1.53 $\times$ 10$^{-3}$ & 1023 & 0.176 & 3\\

\hline
\label{table1}

\end{tabular}
\caption{\footnotesize Input and output parameters for each model in this study. The Reynolds number, Re, based on an rms velocity is given by $\mathrm{Re}=2\,\mathrm{Ro}/\mathrm{Ek}$.  For comparison purposes, we note that the flux Rayleigh number, Ra$_\mathrm{F}$, listed here is lower than those quoted in FH15.  Those values were tabulated incorrectly by a factor of $\pi$.}
\end{table}
}

\shorttitle{Emergence of Supergranulation}
\shortauthors{Featherstone \& Hindman}

\begin{abstract}
We investigate how rotationally-constrained, deep convection might give rise to supergranulation, the largest distinct spatial scale of convection observed in the solar photosphere.  While supergranulation is only weakly influenced by rotation, larger spatial scales of convection sample the deep convection zone and are presumably rotationally influenced.  We present numerical results from a series of nonlinear, 3-D simulations of rotating convection and examine the velocity power distribution realized under a range of Rossby numbers.  When rotation is present, the convective power distribution possesses a pronounced peak, at characteristic wavenumber $\ell_\mathrm{peak}$, whose value increases as the Rossby number is decreased.  This distribution of power contrasts with that realized in non-rotating convection, where power increases monotonically from high to low wavenumbers.  We find that spatial scales smaller than $\ell_\mathrm{peak}$ behave in analogy to non-rotating convection.  Spatial scales larger than $\ell_\mathrm{peak}$ are rotationally-constrained and possess substantially reduced power relative to the non-rotating system.  We argue that the supergranular scale emerges due to a suppression of power on spatial scales larger than $\ell\approx100$ owing to the presence of deep, rotationally-constrained convection.  Supergranulation thus represents the largest non-rotationally-constrained mode of solar convection.   We conclude that the characteristic spatial scale of supergranulation bounds that of the deep convective motions from above, making supergranulation an indirect measure of the deep-seated dynamics at work in the solar dynamo.  Using the spatial scale of supergranulation in conjunction with our numerical results, we estimate an upper bound of 10 m s$^{-1}$ for the Sun's bulk rms convective velocity.
\end{abstract}

\begin{document}
\title{The Emergence of Solar Supergranulation as a Natural Consequence of Rotationally-Constrained Interior Convection}

\author{Nicholas A. Featherstone}
\affil{Department of Applied Mathematics, University of Colorado, Boulder, CO 80309-0526}
\email{feathern@colorado.edu}
\author{Bradley W. Hindman}
\affil{JILA \& Department of Astrophysical and Planetary Sciences, University of Colorado, Boulder, CO 80309-0440}
\maketitle

\section{Introduction}

Supergranulation manifests as the largest distinct scale of convection observed in the solar photosphere.  This spatial scale of convection, approximately 30 Mm in horizontal extent, possesses a clear spectral peak in photospheric Dopplergram power around spherical harmonic degree $\ell\approx120$ (e.g., Hathaway et al. 2000, 2015).  Whereas the Dopplergram power associated with radial motion tends to be dominated by small-scale granulation ($\sim 1$ Mm in size), supergranular flows tend to dominate the horizontal-flow contribution to the spectrum (e.g., Lawrence et al. 1999).  The flow patterns associated with supergranulation were first identified by Hart (1956) and then later associated with cellular convection by  Leighton et al. (1962), but a convincing account of why supergranular scales might be preferentially driven remains elusive.

Simon \& Leighton (1964) suggested that {He~II} recombination could act as a driver of supergranular scales, but in modern numerical simulations of solar convection, ionization effects have failed to reproduce an enhancement of those scales (Stein et al. 2006; Ustyugov 2010). In fact, recent numerical simulations have found the converse; ionization tends to suppress spatial motions at particular horizontal scales (Lord et al. 2014). Other researchers have suggested that self-organization of granules could spontaneously generate larger-scale motions (e.g., Rieutord et al. 2000; Rast 2003; Crouch et al. 2007). Numerical simulations of granulation, however, have yet to self-consistently develop supergranular flows (Stein et al. 2009; Ustyugov 2010; Lord et al. 2014). For an in depth summary of theoretical efforts to explain the origin of supergranulation, we direct the reader to Rast (2003) and Rieutord \& Rincon (2010).

The focus of this letter instead derives in part from an alternative view of supergranulation recently suggested by Lord et al. (2014).  Those authors question whether supergranular scales of convection are actually preferentially \textit{enhanced}.  Instead, they posit that spatial scales larger than supergranulation are in fact preferentially \textit{suppressed}, ultimately resulting in the emergence of supergranulation as a distinct convective scale.  

While Lord et al. (2014) stopped short of identifying a mechanism responsible for the suppression of large-scale power, their question is worth considering.  Horizontal scales of convection larger than supergranulation will almost certainly extend into the deeper convection zone.  Thus, if supergranulation emerges in response to the suppression of large-scale convective motions, it must be an indirect indicator of the deep-seated dynamics at work in the solar dynamo; those dynamics remain difficult to image helioseismically.

\subsection{Rotational Implications}
We suggest that \textit{strong rotational influence}, an effect neglected in earlier, primarily photospheric, investigations of supergranulation, provides a natural means by which deep-seated, large-scale convective motions might be suppressed.  The helioseismic results of Hanasoge et al. (2012), as well as inconsistencies between some numerical results and helioseismic rotation profiles (Featherstone \& Miesch 2015)  imply that convective velocity amplitudes on global scales may be weaker, and thus more rotationally-constrained, than previously believed (see also Miesch et al. 2012 and discussion therein).  These recent results, along with the persistent puzzle surrounding supergranulation's origin, motivate us to consider the consequences of solar convection subjected to varying degrees of rotational constraint.  In this letter, our purpose is twofold:
\begin{enumerate}
\item We outline a suite of numerical convection simulations, designed to illustrate how the convective velocity spectrum responds to the degree of rotational constraint (quantified through a Rossby number).

\item We argue that supergranulation arises as the natural consequence of the intersection between deep, rotationally-constrained convection and rapid near-surface motions that sense rotation only weakly.  Building on this premise, we estimate an upper bound for the solar Rossby number.

\end{enumerate}

\section{The Numerical Experiment}
We have equilibrated a series of nonlinear, rotating convection models in 3-D spherical geometry using the \textit{Rayleigh} convection code. \textit{Rayleigh} solves the anelastic equations using the spectral transform approach described in Glatzmaier (1984). In this approach, system variables are represented by truncated expansions on the sphere using spherical harmonics and, in radius, using Chebyshev polynomials.  See Glatzmaier (1984) as well as Featherstone \& Hindman (2016; hereafter FH16) for additional details concerning the numerical technique and its accuracy properties. \textit{Rayleigh} is being developed through the Computational Infrastructure for Geodynamics (CIG), an NSF-sponsored initiative designed to foster the development and use of open-source codes within the geophysics community.

Our models are purely hydrodynamic in nature and were constructed in similar fashion to the non-rotating models presented in FH16.  We model the innermost 3 density scale heights of the solar convection zone, corresponding to a shell aspect ratio of $\chi \equiv r_\mathrm{inner}/r_\mathrm{outer}$ = 0.759 and dimensional shell thickness of 157 Mm.  Our models are fully-characterized by three nondimensional parameters: a Rayleigh number, a Prandtl number, and an Ekman number.  

As discussed in FH16, a flux Rayleigh number Ra$_\mathrm{F}$ appropriate for this system may be defined as
\begin{equation}
\label{eq:ranum}
\mathrm{Ra}_\mathrm{F} = \frac{\widetilde{g}\widetilde{F}H^4}{c_p\widetilde{\rho}\widetilde{T}\nu\kappa^2},
\end{equation}
where tildes indicate volume averages over the full shell, making Ra$_\mathrm{F}$ a bulk Rayleigh number.  In this expression, $g$ is the gravity, $F$ is the thermal energy flux imposed by radiative heating, and $c_p$ is the specific heat at constant pressure.  $T$ is the temperature, $\rho$ is the density, $\nu$ is the viscosity, and $\kappa$ is the thermal diffusivity.  Both $\nu$ and $\kappa$ are taken to be constant in this study.  For the length scale \textit{H}, we choose the shell depth.  We adopt a Prandlt number Pr = $\nu/\kappa$ of unity in all cases to avoid the onset of unphysical modes known to otherwise arise in rotating anelastic systems (Calkins et al. 2015). The Ekman number Ek is given by
\begin{equation}
\label{eq:Ekman}
\mathrm{Ek} = \frac{\nu}{\Omega H^2},
\end{equation}
where $\Omega$ is the angular frequency of the rotating frame.
\tableone

All models were initiated from non-rotating simulations originally presented in FH16 and were run using 128 Chebyshev collocation points in radius, corresponding to 85 de-aliased Chebyshev modes.  Following equilibration, each model was further evolved for a minimum of one viscous diffusion time.  Our simulations span a range of values for Ra$_\mathrm{F}$ and Ek.  Input parameters for each model, including $\ell_\mathrm{max}$, the maximum spherical harmonic degree employed in the truncated expansion, are provided in Table 1.

\subsection{Results}\label{sec:results}
That rotation induces fundamentally different dynamics than is otherwise achieved in its absence is hardly a novel result (see e.g., Busse 2002), but  it is nonetheless worth illustrating here.  Figure \ref{fig:composite}$a$ depicts a snapshot of the radial velocity pattern realized in a non-rotating case run at Ra$_\mathrm{F}$ = $6.81\times10^6$.  The snapshot is sampled at a radius $r/r_\mathrm{outer}= 0.99$ or equivalently, 5.6 Mm below the outer boundary.  Convective cells are distributed isotropically across the spherical surface and manifest on spatial scales ranging from the box-scale (i.e., the dipolar mode) down to the dissipation scale (see discussion in FH16). 

\figone

Figure \ref{fig:composite}\textit{b} illustrates the same system with rotation (Ek = $1.91\times10^{-4}$).  Convection in the equatorial region of this system possesses the columnar roll patterns that commonly arise in rotating convection (often termed ``banana-cells" or ``Busse columns").  Convective patterns in the polar regions are also columnar, but they appear considerably more complex due to their intersection with the spherical surface. 

These non-rotating and rotating flow patterns possess very different horizontal velocity power spectra as shown in Figure \ref{fig:composite}\textit{c}.  There, we plot $P_\ell$, the normalized power spectra associated with the horizontal velocity field, given by 
\begin{equation}
P_\ell \equiv  \frac{1}{\widetilde{P}}\sum\limits_{m=-\ell}^{\ell}\mathbf{v}_{\ell m}\cdot\mathbf{v}_{\ell m}^{*},
\end{equation}
where the normalization constant $\widetilde{P}$ is the total integrated power,
\begin{equation}
\widetilde{P} = {\sum\limits_{\ell=0}^{\ell_\mathrm{max}}\sum\limits_{m=-\ell}^{\ell}\mathbf{v}_{\ell m}\cdot\mathbf{v}_{\ell m}^{*}},
\end{equation}
and where $\mathbf{v}_{\ell m}$ is the spherical harmonic transform of the horizontal velocity vector.  All spectra were calculated at radius $r/r_\mathrm{outer}= 0.99$ and temporally averaged over one third of a viscous diffusion time.  The spectra were normalized in order to emphasize the differences in shape between the non-rotating system (blue curve) and the rotating system (black curve).  For the non-rotating case, power increases more or less uniformly from small to large scales as is typical of non-rotating convection (e.g., Ahlers et al. 2009; FH16).  The spectral distribution of velocity power in the rotating regime is somewhat more complex.  For the rotating case, in addition to total horizontal velocity power, we also plot the contributions from axisymmetric motions (green; i.e., differential rotation and meridional circulation) and from the non-axisymmetric motions (red; i.e., convection).

Axisymmetric motions are the primary contributor to low-wavenumber power in the rotating case and exhibit strong, even $\ell$-parity as is expected for the hemispherically symmetric differential rotation and meridional circulation.  Convective power increases monotonically as a function of $\ell$ until some critical value $\ell_\mathrm{peak}$ (equal to 77 in this example) is reached.  Above $\ell_\mathrm{peak}$, power decreases monotonically in a fashion resembling that of the non-rotating system.

The similarity of the rotating and non-rotating velocity spectra for $\ell\ge\ell_\mathrm{peak}$ suggests that those spatial scales of convection are only weakly influenced by rotation.  We can explore this interpretation quantitatively by examining how $\ell_\mathrm{peak}$ varies with a bulk Rossby number, Ro, given by 
\begin{equation}
\label{eq:rossby}
\mathrm{Ro} = \frac{\widetilde{U}}{2\Omega H}.
\end{equation}
Here $\widetilde{U}$ is a typical velocity amplitude and $H$ is, again, the shell depth.  We adopt the rms convective velocity amplitude for $\widetilde{U}$, taking the mean over the full volume of the shell. The Rossby number expresses the ratio of the rotation period to a typical convective overturning time. Low-Ro convection is heavily influenced by the Coriolis force, which otherwise plays a minimal role in high-Ro convection.  The Rossby number and $\ell_\mathrm{peak}$ are provided for each of our models in Table 1.

\figtwo

A series of convective velocity spectra from our suite of models, spanning a range of Rossby numbers, are plotted in Figure \ref{fig:progression}. In order to clearly illustrate that the location of peak spectral power, $\ell_\mathrm{peak}$, is dependent on Ro, these spectra have been normalized to a peak value of unity.  We note that the violet curve, which corresponds to a relatively high value of Ro, exhibits anti-solar differential rotation (e.g., Gastine et al. 2014).  This phenomenon, characterized by retrograde rotation in the equatorial region and prograde rotation in the polar regions, occurs in regimes of weak rotational influence. Convection in this regime does not evince the columnar patterns of Figure \ref{fig:composite}{b}, but instead possesses a convective velocity spectrum similar to that of non-rotating convection (black curve).

We have plotted $\ell_\mathrm{peak}$ against Ro$^{-1}$  for each of our models in Figure \ref{fig:progression}\textit{b}.  Symbols for solutions that display anti-solar behavior are colored blue to distinguish them from rotationally-constrained solutions (red symbols).  The variation of $\ell_{peak}$ for the four lowest-Ro cases seems to follow a scaling law $\ell_\mathrm{peak} \sim \mathrm{Ro}^{-1/2}$ (solid reference line; not a fit).   Such a scaling occurs when Coriolis, buoyancy, and inertial forces are of comparable amplitude in the momentum equation (see e.g., Stevenson 1979; Ingersoll \& Pollard 1982).   This scaling law may be derived by equating the Coriolis and inertial terms in the $\mathbf{\hat{z}}$-vorticity equation, where $\mathbf{\hat{z}}$ is a unit vector parallel to the rotation axis. Doing so yields
\begin{equation}
\label{eq:CIA}
\Omega\frac{\partial u_z}{\partial z} \sim \mathbf{u}\cdot\mathbf{\nabla}\omega_z,
\end{equation}
where $\omega_z$ is the z-component of vorticity, and where we have assumed incompressibility to simplify the analysis.  If we consider a convective column in the polar region with axial length $H$ and horizontal diameter $L = 2\pi r_\mathrm{outer}/\ell_\mathrm{peak}$, we find
\begin{equation}
\label{eq:CIA2}
\Omega\frac{\widetilde{U}}{H}\sim \frac{\widetilde{U}^2}{L^2},
\end{equation}
which may be rearranged to yield
\begin{equation}
\frac{L^2}{H^2} \sim \frac{\widetilde{U}}{\Omega H} = 2\mathrm{Ro}.
\end{equation}
Noting that $H=r_\mathrm{outer}(1-\chi)$, we see that
\begin{equation}
\frac{L}{H} = \frac{2\pi}{(1-\chi)}\frac{1}{\ell_\mathrm{peak}}\sim \mathrm{Ro}^{1/2}.
\end{equation}

A similar analysis equating the Coriolis and viscous forces would alternatively yield the $\mathrm{Ek}^{1/3}$ scaling expected near convective onset (Chandrasekhar 1953; something we do not observe in this data set).  As a result, we conclude that diffusion does not play a dominant role in our low-Ro models.  Moreover, the location of $\ell_\mathrm{peak}$ in our spectra represents a break between those spatial scales where the Coriolis force is dominant ($\ell < \ell_\mathrm{peak}$) and those where inertia dominates ($\ell > \ell_\mathrm{peak}$).

Finally, we note that the transition between solar and anti-solar behavior corresponds to a Rossby number of 0.09 as defined here, and the associated $\ell_{peak}\approx20$ loosely corresponds to depth of the convective layer.  Thus, \textit{if} the Sun possesses giant cells in the traditional sense (i.e. convection on horizontal scales comparable to the convection zone depth), the convection zone is only marginally rotationally-constrained.  We suggest that this is not the case.

\section{Perspectives on Supergranulation \\ \& Global Dynamics}

Supergranulation is relatively insensitive to rotational influence; a naive estimate of its Rossby number, adopting $H=30$ Mm and $\widetilde{U}=300$ m s$^{-1}$, yields $\mathrm{Ro}=2$.  This fact is most likely why earlier investigations into supergranulation's origin have neglected to consider the effects of rotation.  Nevertheless, supergranulation manifests in a region of the convection zone where rotational influence is clearly transitioning with depth.  


The deep convection zone, for instance, must be rotationally constrained to \textit{some} extent by virtue of its prograde equatorial differential rotation (Glatzmaier \& Gilman 1982; Brun \& Toomre 2002; Gastine et al. 2014, Guerrero et al. 2013, K\"{a}pyl\"{a} et al. 2014, Featherstone \& Miesch 2015).  The near-surface-shear layer in the upper convection zone (e.g., Howe 2009) is in turn thought to be maintained via inward angular momentum transport by high-Ro convective motions (e.g., Foukal \& Jokipii 1975; Gilman \& Foukal 1979; Hotta et al. 2015).  

This Rossby-number transition was verified helioseismically by Greer et al. (2016), who measured a transition from Ro$\,\approx\,$5 to Ro$\,\approx\,$0.2 across the upper 10 Mm of the convection zone.  Presumably, Ro is even lower below the base of the near-surface-shear layer.  We suggest that the transition to low-Rossby-number convection at depth, in combination with the spectral behavior described in \S\ref{sec:results}, ultimately lead to the emergence of a supergranular scale.

\subsection{A Conceptual View of the Photospheric Velocity Spectrum}
To illustrate how supergranulation might emerge through a transition from high-Ro to low-Ro convection, we consider the photospheric velocity power associated with \textit{horizontal flows}.  We thus neglect radial motion associated with photospheric driving at the granular scale, a dominant feature in Dopplergram power that cannot be captured in our models which lack radiative transfer.  We focus instead on deep-seated convection, ultimately generated in response to that photospheric driving, and whose overturning motion is primarily horizontal in the photosphere (as is the case with supergranulation).  We consider a two-component flow.  One component is associated with near-surface, high-Ro motions, whose streamlines and power contributions we sketch using red in the schematics shown in Figure \ref{fig:cartoon}.  The other component, indicated in blue, is associated with motions that extend into the deep convection zone.  We consider two cases: non-rotating convection and rotating convection subject to a transition in Rossby number with depth, as is the situation in the Sun. 


\figthree

In the absence of any rotation (Figure \ref{fig:cartoon}$a$), bulk motions with a horizontal scale comparable to the convection zone depth (i.e., giant cells) would be excited in analogy to the box mode of classical Rayleigh B\'{e}nard convection (e.g., Ahlers et al. 2009).  Superimposed upon those deep-seated motions would be shallow near-surface motions (i.e., granulation and supergranulation).  The photospheric velocity spectrum would reflect this wide range of horizontal convective scales, and the horizontal velocity power would naturally increase from small scales up to giant-cell scales (Figure. \ref{fig:cartoon}$b$).

In the rotating case (Figure \ref{fig:cartoon}$c$), the spectrum of the near-surface, high-Ro motions will remain largely unchanged. The deep convective motions will manifest on the giant-cell scale \textit{only} if the deep convection zone is in a state of marginal rotational constraint ($\mathrm{Ro}\approx0.09$; see \S\ref{sec:results}).  In that case, the spectrum of Figure \ref{fig:cartoon}$b$ will remain largely unchanged.  

If the Sun does not reside at a marginally stable point in parameter space, its convection will assume a thin, columnar structure whose characteristic horizontal length scale is Ro-dependent.  The horizontal velocity spectrum would then peak at a higher wavenumber than in the non-rotating case, as illustrated in  Figure \ref{fig:cartoon}$d$, where we have chosen to illustrate the situation where deep convective pwoer peaks at a similar wavenumber to that of near-surface power.  We note that the relative location of those two peaks depends on both the structure of the Ro-transition region and the Rossby number of the deep convection.

This spectral behavior, resulting from rotational influence on the deep convective motions, provides a natural explanation for the emergence of a supergranular scale of convection and the apparent absence of giant-cells in the traditionally expected sense.  We emphasize that Figure \ref{fig:cartoon} is only schematic in nature.  In the high-Re, low-Ro limit, these columns would lose coherency and break up into even smaller-scale, yet still rotationally constrained, motions (e.g., Sprague et al. 2006).  Similarly, the near-surface cellular motions that we have drawn represent a much more complicated dynamic involving the coalescence of descending plumes that penetrate into regions of broad, upwelling flow (e.g., Stein \& Nordlund 1998).

\subsection{Implications for Supergranulation}

The Sun is rotating, two independent lines of helioseismic evidence indicate that solar convection is transitioning from high-Ro behavior at the surface, to low-Ro behavior at depth, and convective power on large-scales is naturally suppressed in low-Ro regimes (Hanasoge et al. 2012; Greer et al. 2016; \S\ref{sec:results}).  Based on these facts, we conclude that the lack of significant observed power below $\ell\approx120$, typically associated with supergranulation, is most easily explained by the presence of deep-seated, rotationally-constrained convection.  A more precise statement is difficult without knowing if and how giant cells contribute to the breadth of the supergranular peak in photospheric power, which evinces significant power in the range ($80 \lesssim \ell \lesssim120$).

In essence, supergranulation emerges as the largest distinct scale of convection observed at the solar surface because larger convective motions are suppressed due to low-Ro dynamics associated with the deep convection zone.  Those scales of convection are behaving in analogy to the rotationally-dominated band of wavenumbers in Figure \ref{fig:composite}$c$.  Note the powerful implication of this conclusion: the spatial size of supergranulation provides \textit{an upper bound} to that of the deep-seated, rotationally constrained motions.  

 As a result, we can estimate an upper bound for the solar Rossby number, as defined in Equation \ref{eq:rossby}, by using the length scale of supergranulation, along with our numerical results.  We see from Figure \ref{fig:progression}$b$ than $\ell_\mathrm{peak}\approx100$ corresponds to $\mathrm{Ro}\approx0.01$.  Using the depth of the solar convection zone for $H$, and the solar rotation rate for $\Omega$, yields an rms $\widetilde{U}$ for the solar convection zone of 10 m s$^{-1}$.  This estimate is not dissimilar from the helioseismic bounds established by Hanasoge et al. (2012) and the theoretical estimates of Miesch et al. (2012), though we note that our estimate represents a volumetric average of the velocity amplitude.  Further refinement of our estimate will involve a systematic investigation of the role played by magnetism and additional density stratification.  

\subsection{Implications for Giant Cells}
Finally, if the deep convective motions were only weakly rotationally constrained, we would expect a peak of power at the giant-cell scale ($10 \lesssim \ell \lesssim20$).  As we do not see such a peak (e.g., Hathaway et al. 2000, 2015; Lord et al. 2014), and observe only surprisingly weak flows at those scales instead (Hathaway et al. 2013), we conclude that giant-cells in the traditional sense do not exist.  This is further evidence that the Rossby number of the deep motions is low, and that their velocity power peaks at a scale \textit{smaller than or comparable to} that associated with supergranulation. Moreover, low-Ro conditions in the bulk of the convection zone may explain why only weak large-scale organization has been observed in the supergranular pattern (Lisle et al. 2004; Hathaway et al. 2013).

\section{Conclusions}
We close with a restatement of the principle conclusions of this letter.
\begin{enumerate}
\item Low-Rossby-number solar convection provides a natural mechanism by which large-scale convective power is suppressed and may explain the lack of power for spatial scales larger than supergranulation.  Such an absence of large-scale power explains the difficulty in unambiguous detection of giant cells, which have been expected to manifest at wavenumbers of $10 \lesssim \ell \lesssim20$.  
\item As a corollary, the supergranular scale represents a lower bound on $\ell_\mathrm{peak}$ in the solar convection zone, thus providing an implicit upper bound for the solar Rossby number.  We estimate the solar Rossby number to be of order 10$^{-2}$, which yields an rms convective velocity amplitude for the Sun of order 10 m s$^{-1}$.
\end{enumerate}

\acknowledgments
We thank Keith Julien for his help interpretting our peak wavenumber scaling, and we thank Mark Rast for several useful discussions.  We further thank Ben Brown and an anonymous referee for several thoughtful comments on the initial draft of this letter. This work was supported by NASA grants NNX14AC05G (Featherstone \& Hindman) and NNX14AG05G (Hindman).  Featherstone and the development of $Rayleigh$ were primarily supported by the Computational Infrastructure for Geodynamics (CIG), which is supported by the National Science Foundation award NSF-094946.  This work used computational resources provided by an award of computer time on the Mira supercomputer through the Innovative and Novel Computational Impact on Theory and Experiment (INCITE) program.  This research used resources at ALCF, which is a DOE Office of Science User Facility supported under Contract DE-AC02-06CH11357. Further computational resources were provided through NASA HEC support on the Pleiades supercomputer. 


\end{document}